\documentclass{article}
\usepackage{graphicx}
\usepackage{anyfontsize}
\graphicspath{ {./draft_jpeg/} }
\begin{document}

\newcommand{\D}{\ensuremath\mathrm{d}}
\begin{center}
{\large \bf Chemical characterization of dislocation in yttria-stabilized zirconia}
\\
{Kyoung-Won Park$^{a,b,*}$}

{\small $^{a}$ Center for Biomaterials, Korea Institute of Science and Technology, Seoul, 02792, Republic of Korea}\\
{\small $^{b}$ Department of Materials Science and Engineering, Massachusetts Institute of Technology, Cambridge, Massachusetts 02139, USA}\\

\end{center}

\begin{abstract}

This study demonstrates that a space charge layer is formed on dislocation during mechanical deformation at elevated temperature. High density of dislocation lines is generated in bulk single crystalline Y$_{2}$O$_{3}$ stabilized ZrO$_{2}$ (YSZ) by uniaxial compression at elevated temperature. The creation of dislocation is proven with transmission electron microscopy (TEM). Then, energy-dispersive X-ray spectroscopy (EDS) and electron energy loss spectroscopy (EELS) are used to explore the changes in the composition \textit{on} and \textit{away from} the dislocation lines. Also, it is clarified that segregation of dopant atoms (yttrium) on the dislocation line is induced by high temperature annealing. Comparing the compositional variations with and without thermal annealing, we study the space charge layer formed on dislocation lines in a doped system.
\\
\end{abstract}

{\textbf{keywords}}: Y$_{2}$O$_{3}$ stabilized ZrO$_{2}$ (YSZ); uniaxial compression; dislocation; segregation of dopant; space charge layer\\

\vfill

Corresponding Author\\
$^{*}$E-mail: exclaim27@kist.re.kr\\

\newpage

\section{Introduction}
Structural defects in materials are of scientific interest because unique properties of materials are determined by type, density, and distribution of the defects. In oxygen ion conductors, the concentration of mobile oxygen vacancy which takes part in a real oxygen ion transport process, is the crucial factor in electrical conductivity in a single crystal. For insertion of a high concentration of oxygen vacancy, homogeneous chemical doping is commonly used. For example, acceptor dopants in representative ion conductors such as CeO$_{2}$ and ZrO$_{2}$-based systems greatly increase ionic conductivities [1-3]. On the other hand, localized doping of charged structural defects is another way to modify ionic conduction. Polycrystalline CeO$_{2}$ and ZrO$_{2}$-based materials are known to contain positively charged grain boundaries. Because of the existence of the positively charged plane defect, charge carriers redistribute themselves along the defects due to electrostatic force between the charged defects and charge carriers, resulting in reduced ionic conduction and enhanced electronic conduction [4]. The origin of the inherent resistance of grain boundary in ionic conduction has been well explained using a space charge model [4-10]. However, how dislocation affects electrical properties of metal oxides and why the dislocation shows the characteristic conduction behavior has not yet been understood.

Since an one-dimensional line defect, \textit{i.e.}, dislocation, can accommodate oxygen vacancies on the core, it is expected to guide fast ionic conduction in metal oxide materials. Yet, very few experimental studies have found that the existence of dislocations increases electrical conductivity [11-12]. Otsuka et al.[11] showed that a plastically deformed YSZ crystal has higher conductivity by $<$ 10 $\%$ than an undeformed one. Moreover, it was found that the electrical conductivity increased with increasing the amount of plastic deformation [11]. Nevertheless, they concluded without any experimental evidence that the enhanced electrical conductivity contributed to the increased ionic conductivity. Furthermore, the fundamental issue of how the dislocations could improve the ionic transport was never thought to be related to the chemical compositional variation. Recently, a pioneering work gave better understanding of the nature of the increased ionic conductivity in the presence of dislocations using a space charge model [12]. But, that work also failed to provide direct evidence, which would have supported the space charge model, of compositional modulation near dislocations. I consider it essential that further work ascertains whether chemical modulation around dislocations determines ion-conducting behavior. 

Many recent studies on the role of dislocations in the ionic conductivity have been focused on the misfit dislocations existing in the heterointerface between two different materials in order to readily generate dislocations in strained samples by lattice mismatch [13-15]. For all that, it is difficult to explore the sole effect of dislocations on the variation of ionic conductivity, that is, whether the characteristic conductivity is indeed caused by misfit dislocations. In addition, high density of dislocations at the heterointerface might induce the dislocations’ relaxation at the grain boundary, not at the heterointerface. Therefore, due to this complexity, a way to introduce only dislocations into a simple system should be suggested.

Mechanical deformation is a common way to generate dislocations by applying enough mechanical energy to cause plastic deformation in the sample. But in the case of brittle materials such as ceramics and semiconductors, the density of the dislocations introduced by plastic deformation is limited because the samples easily undergo fracturing. High-temperature compression is one of the effective methods to introduce dislocations into a brittle crystalline solid.

In this study, I conduct uniaxial compression on bulk single crystalline YSZ at elevated temperature to create dislocations and elucidate compositional variation occurring near the dislocations. Also, post-heat treatment is conducted to clarify the dopant segregation phenomenon that can happen in a doped ionic conducting system. On the basis of the present findings, I address space charge layer formation on dislocations in YSZ in view of the redistribution of oxygen and yttrium atoms.

\section{Model system}

Fluorite oxides are the classical oxygen ion-conducting materials. This fluorite structure is able to sustain a high degree of substitution and allow large volume expansion during a reduction process [16-17]. This means fluorite-structured systems have very stable structures which can maintain themselves even under external stimuli [18]. Therefore, I expect the fluorite structure would persist even with the formation of high density of dislocations. Among fluorite systems, YSZ is considered to be a proper structural system for conducting this research, because the host cation, Zr$^{4+}$, has a fixed host cation oxidation state. This characteristic provides a simpler characterization of the chemistry of dislocations than other reducible/oxidizable host cations. Also, the readily obtainable bulk single crystalline substrate is another benefit of YSZ. This is because creating dislocations in a single crystal is better for clarifying the sole effect of dislocations on conductivity without the consideration of any interface effect occurring between thin film and substrate.

\section{Methods}

Bulk single crystalline 8 mol.$\%$ YSZ (100) ($5 \times 5 \times 2$ mm$^{3}$) used in this study was bought from MTI Corporation. To generate dislocations in the sample, uniaxial compression tests (MTS 810 Materials Testing System) were carried out on the YSZ sample at a constant strain rate of 10$^{-4}$/s at various temperatures, \textit{i.e.}, room temperature, 300, 500, 650 and 800 $^{o}$C. 

A TEM sample was prepared by using focused ion beam (FIB, FEI Helios 600 NanoLab). To eliminate any charge effect and protect the sample surface from Ga$^{+}$ ion beam damage, Pt/Pd (80/20) of 30 nm was deposited on the compressed sample by using a sputter coater (EMS 300T D). To protect the sample surface from the ion source, Pt was additionally deposited by 500 nm with electron beam and 2 $\mu$m with ion beam with gas injection system (GIS). The sample was Ga$^{+}$ ion milled at 30kV, various current. A tungsten (W) omniprobe was used to pick up the milled plate and attach it on a Cu half TEM grid. After the sample attachment on the grid, fine milling was conducted at low currents by reducing the milling current gradually. Afterwards, fine milling at reduced ion energy (Ar source, 500 eV) was carried out to remove the surface amorphous layer (Fischione NanoMill 1040).

The distribution and density of the dislocations were investigated using conventional transmission electron microscope (FEI TECNAI G2) at accelerating voltage of 120 kV. For chemical characterization, a scanning TEM (STEM, JEOL 2010F) imaging was performed. Here, a 200 kV electron beam is focused down to a spot with probe size of 0.7 nm, and scanned across a thinned sample. Composition changes \textit{on} and \textit{away from} dislocations were analyzed using energy-dispersive X-ray spectroscopy (EDS) and electron energy loss spectroscopy (EELS).

\section{Results and Discussion}
Figure 1 shows the representative engineering stress-strain curves recorded during uniaxial compression at room temperature and 800 $^{o}$C. As expected, single crystalline YSZ is very brittle at room temperature, while it shows enhanced plastic strain ($\approx$ 4.5$\%$) and reduced yield strength (775 to 630 MPa) at 800 $^{o}$C. This enhanced plastic strain provides the indirect evidence of an increased number of dislocations in YSZ. Hence, it is expected that the sample compressed at 800 $^{o}$C would present increased electrical conductivity [11]. 

\includegraphics[width=120mm]{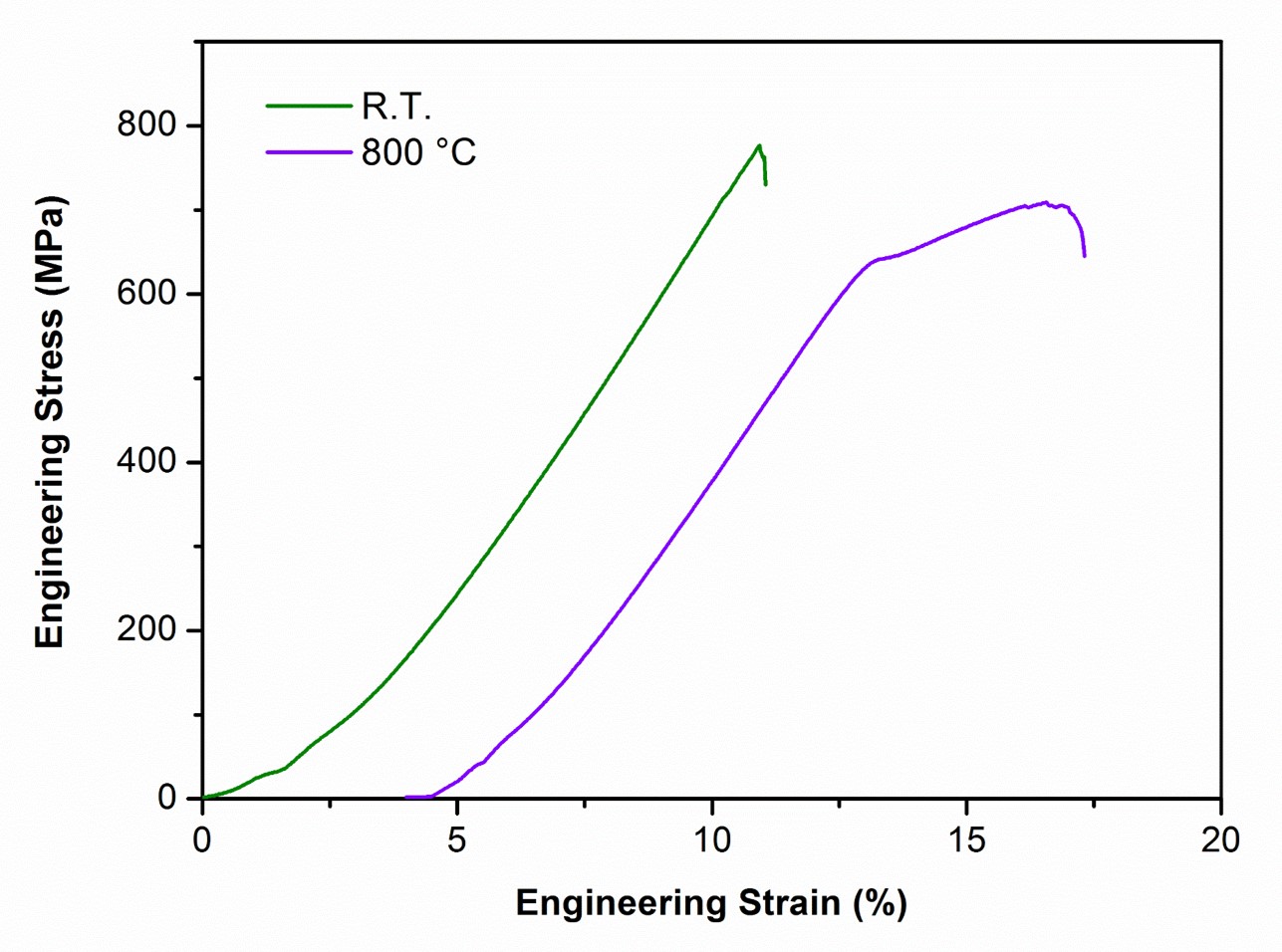}\
 
\textbf{Figure 1} Engineering stress-strain curves of bulk single crystalline YSZ obtained during uniaxial compression at room temperature and 800 $^{o}$C.
\\
\\
Figure 2(b) shows a TEM bright-field (BF) image of the as-compressed YSZ at 800 $^{o}$C after a two beam condition was satisfied on the sample (Figure 2(a)). Many dislocation lines (actually, the strain fields near dislocation cores) are observed. The distribution of dislocations is not homogeneous, but relatively tangled and twisted in localized regions.\

The chemical inhomogeneity around dislocation lines was investigated with Z-contrast images obtained with a scanning TEM high-angle annular dark field (STEM-HAADF), as shown in Figure 2(c). Unlike a STEM-HAADF image of an undeformed YSZ sample (not shown in this paper), the STEM-HAADF image in Figure 2(c) shows bright lines and dots in the whole area of the sample, the contrast of which is considered to originate from chemical inhomogeneity near dislocations formed during mechanical deformation. Considering that the intensity of the high angle scattering basically depends only on the types of constituent atoms, the high Z atoms appear brighter since they scatter more strongly. Therefore, the bright contrast in the STEM-HAADF image in Figure 2(c) is expected to be caused by 1) depletion of light element, \textit{i.e.}, oxygen, and/or by 2) accumulation of a heavy element such as yttrium on dislocation lines rather than in the matrix region. 

\includegraphics[width=130mm]{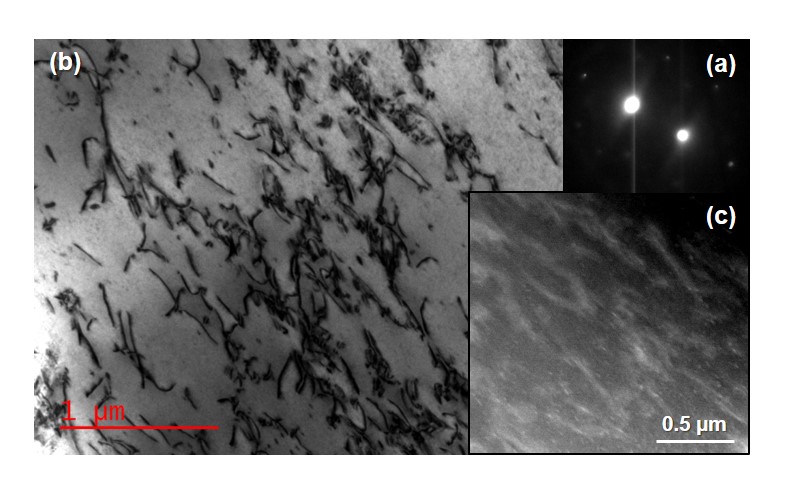}\
 
\textbf{Figure 2} \textbf{(a)} Diffraction pattern collected under a two beam condition. \textbf{(b)} TEM-BF image under a two beam condition showing the distribution of dislocations in the as-compressed YSZ at 800 $^{O}$C. \textbf{(c)} STEM–HAADF image of the as-compressed YSZ at 800 $^{O}$C.
\\
\\
Compositional variation around dislocations is investigated with EDS analysis if there is atomic redistribution around the dislocations during mechanical deformation. For reproducible data, EDS analysis was performed at several points and then the values were averaged. Considering that the spot size of the electron beam used here for STEM-EDS was 0.7 nm, the point analysis of EDS is expected to resolve the chemistry of the dislocation core from that of the matrix with consideration of the beam broadening effect. Figure 3 presents the average atomic compositions \textit{on} dislocations (the bright lines in the STEM-HAADF image) and \textit{away from} dislocations (the dark matrix region in the STEM-HAADF image). Oxygen content largely decreases on dislocations compared to the matrix region, while yttrium (dopant) content is kept almost unchanged. Accordingly, it is suggested that oxygen is depleted on dislocations, interpreting the accumulation of oxygen vacancies on dislocation core. The accumulation of positively charged oxygen vacancies is expected to show a positively charged dislocation line like the grain boundary of YSZ in space charge model [19-20]. However, since the detection rate of light elements such as oxygen with EDS is low, only qualitative change in oxygen content can be identified by the EDS analysis.

\includegraphics[width=130mm]{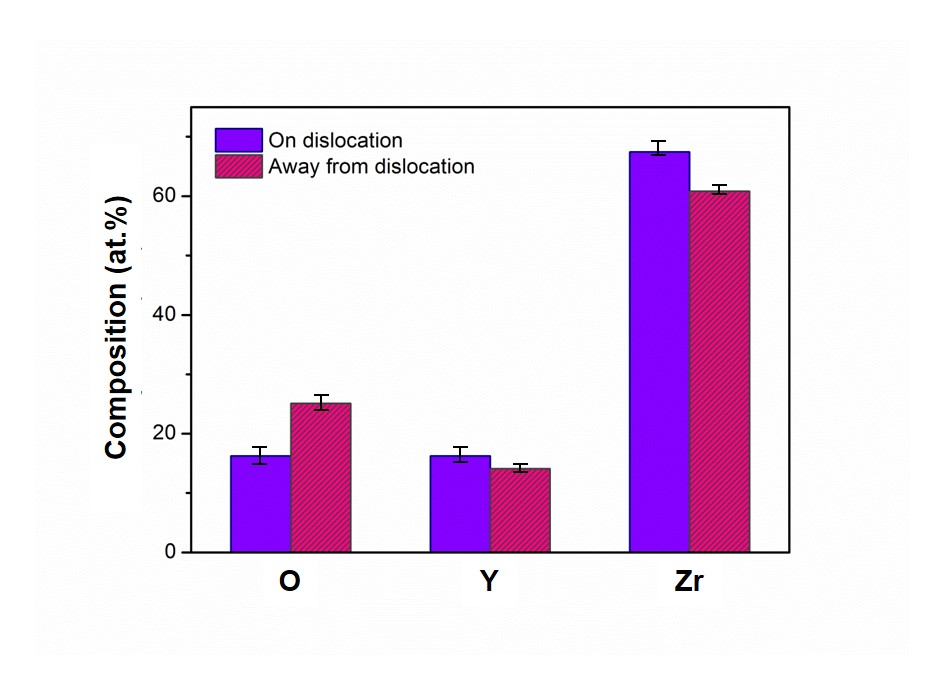}\
 
\textbf{Figure 3} Average atomic compositions \textit{on} dislocation line (bright line in Figure 2(c)) and \textit{away from} dislocation line (dark matrix region in Figure 2(c)) in the as-compressed YSZ by EDS
\\
\\

To overcome the limited detection rate of oxygen with EDS and confirm whether oxygen ions are indeed depleted (or oxygen vacancies are accumulated) on dislocations, an EELS equipped with STEM mode is utilized. The oxygen K-edge EEL spectra obtained from both dislocations and the matrix region are shown in Figure 4, with background signal subtraction. In accordance with the EDS result, the electron count of the oxygen K-edge obtained from the dislocation line is diminished compared to that from the matrix area. These EELS and EDS results indicate oxygen deficiency due to segregation of oxygen vacancies on dislocations, forming a positively charged dislocation core in YSZ. Due to electrostatic interaction between positively charged dislocations and oxygen vacancies, positively charged oxygen vacancies might deplete near the dislocation, while negatively charged carriers are accumulated. The concentration deviation of the oxygen vacancy in the vicinity of dislocations from the bulk value will lead to the formation of a space charge region, as suggested in the space charge model for a grain boundary in YSZ [19-20]. 

\includegraphics[width=130mm]{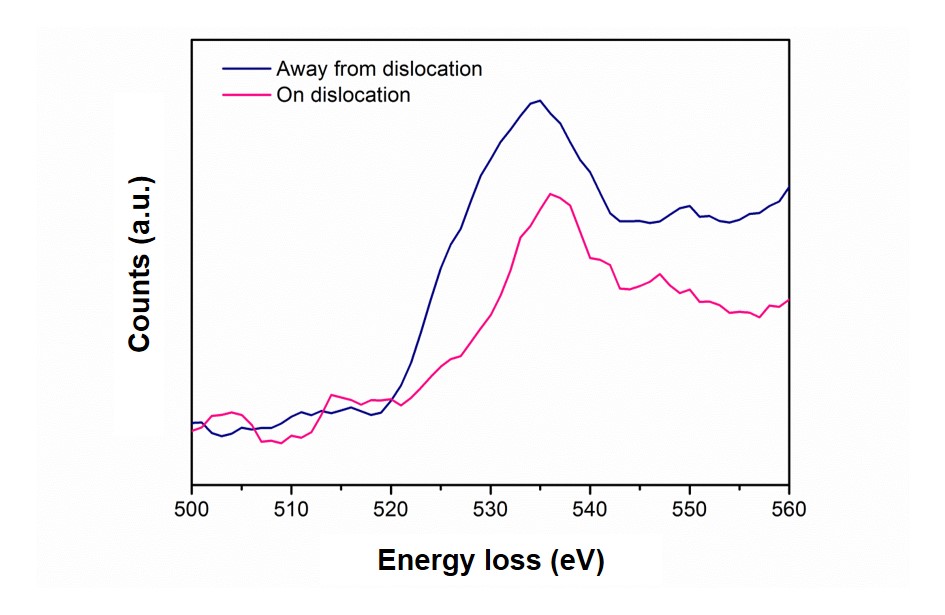}\
 
\textbf{Figure 4} Electron energy loss spectra of oxygen K-edge obtained \textit{on} dislocation line and \textit{away from} dislocation line after background signal subtraction. 
\\
\\
Unfortunately, in this experimental study, only the qualitative oxygen deficiency/oxygen vacancy accumulation on the dislocation cores was proven, although the corresponding decrease of oxygen vacancy concentration near the positively charged core of YSZ was reported in the previous study of hybrid MC-MD simulation [19]. This may be because of limited spatial resolution of STEM-EDS or insufficient quantification ability of TEM to precisely capture the oxygen vacancy depletion in the space charge layer region in the vicinity of dislocation, as insisted in ref. 19. Nevertheless, the chemical characterization of STEM-EDS and -EELS demonstrates oxygen depletion/oxygen vacancy accumulation in the dislocation core region. Accordingly, the space charge layer formed on dislocation would act as a blocking layer in oxygen ionic transport as suggested in the space charge model of the YSZ grain boundary [10].

The bright line of the STEM image in Figure 2(c) is able to come from dopant segregation (yttrium in YSZ), as suggested in the grain boundary in a doped system [21-23]. However, the dopant segregation was not clearly observed in this study (Figure 3); Considering that oxygen diffusion in YSZ (D$_{O}$ = 2.7 -- $3.8 \times 10^{-14}$ m$^{2}$/s at 600 $^{O}$C for bulk sample with grain boundary fraction of 0.26, D$_{O}$ = 5.1 -- $7.2 \times 10^{-14}$ m$^{2}$/s at 600 $^{O}$C with grain boundary fraction of 0.06 [9], D$_{O}$ = $9 \times 10^{-14}$ m$^{2}$/s at 500 $^{O}$C in thin YSZ film [24]) is extremely higher than yttrium diffusion (D$_{Y}$ = $\approx$ 1 -- $3 \times 10^{-20}$ m$^{2}$/s at 1270 -- 1700 $^{O}$C [25]), it is considered that approximately 30 min. taken for quasistatic compression at 800 $^{O}$C is insufficient to make yttrium atoms diffuse to the dislocation core. Consequently, in this study, prominent compositional variation shown in the STEM image (Figure 2(c)) arises only from the depletion of oxygen atoms which are readily movable. Hence, the space charge layer formed on dislocations during mechanical deformation is regarded as showing a similar aspect to that formed in grain boundaries in an undoped system. In general, since grain boundaries in bulk YSZ are generated during sintering process at very high temperature ($>$ 1200 $^{O}$C), the dopant segregation takes place around the grain boundary in the doped system with the help of thermal diffusion of the dopant. Thus, yttrium migration might not occur at the temperature where the compression test was performed in this study (\textit{i.e.}, 800 $^{O}$C), as observed in the EDS result in Figure 3.
Here, the question arises as to whether yttrium (dopant) would be accumulated on the dislocation lines if high enough thermal energy is provided to induce yttrium to diffuse. For investigation of temperature effect on the dopant segregation around dislocations, the as-compressed YSZ sample was annealed at 1200 $^{o}$C for 30 min. The change in the density and distribution of dislocations by post-heat treatment was investigated with TEM-BF imaging under a two beam condition (Figure 5(a)). The chemical inhomogeneity was also observed by STEM-HAADF imaging (Figure 5(b)). It is found that density of dislocation decreases by heating, but the configuration is still twisted. It is interesting that the dislocation contrast in the TEM-BF image is the same as that in the STEM-HAADF image, which indicates a much clearer STEM-HAADF contrast than that in the as-compressed sample. This clearer contrast in the post-annealed sample is related to sharper chemical inhomogeneity around dislocations, which might stems from segregation of the heavy element (yttrium) and the corresponding narrower space charge layer width, consistent with the space charge layer model in a doped system.\\
\\
\includegraphics[width=110mm]{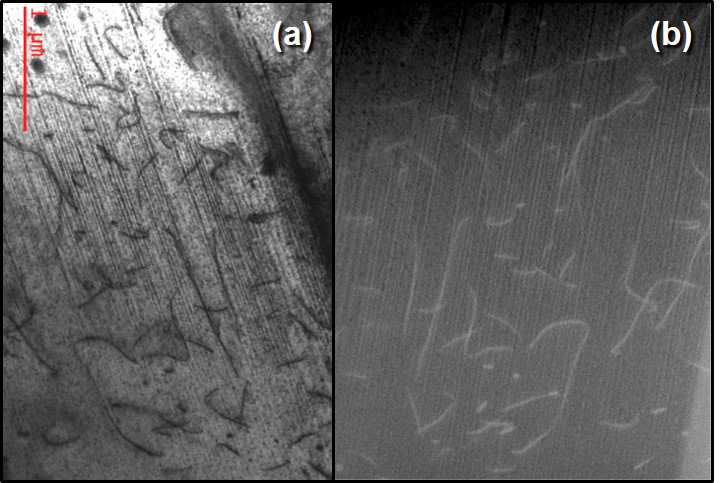}\
 
\textbf{Figure 5} \textbf{(a)} TEM-BF image of the post-annealed YSZ taken under a two beam condition. \textbf{(b)} STEM–HAADF image of the post-annealed YSZ. 
\\
\\
To explore whether the dopant segregation in dislocations is indeed caused by the post-heating, EDS analysis was conducted \textit{on} and \textit{away from} dislocations in the annealed sample (Figure 6). For comparison, the average atomic compositions acquired from the as-compressed and the annealed samples are summarized in Table 1. The tendency of oxygen deficiency on dislocation line is coincident with that in the as-compressed sample. However, the difference in the oxygen content is smaller than in the as-compressed sample. Also, the yttrium segregation becomes more significant in the annealed sample, which is in accordance with the dopant segregation phenomenon observed at the grain boundary formed at higher temperature.

\includegraphics[width=130mm]{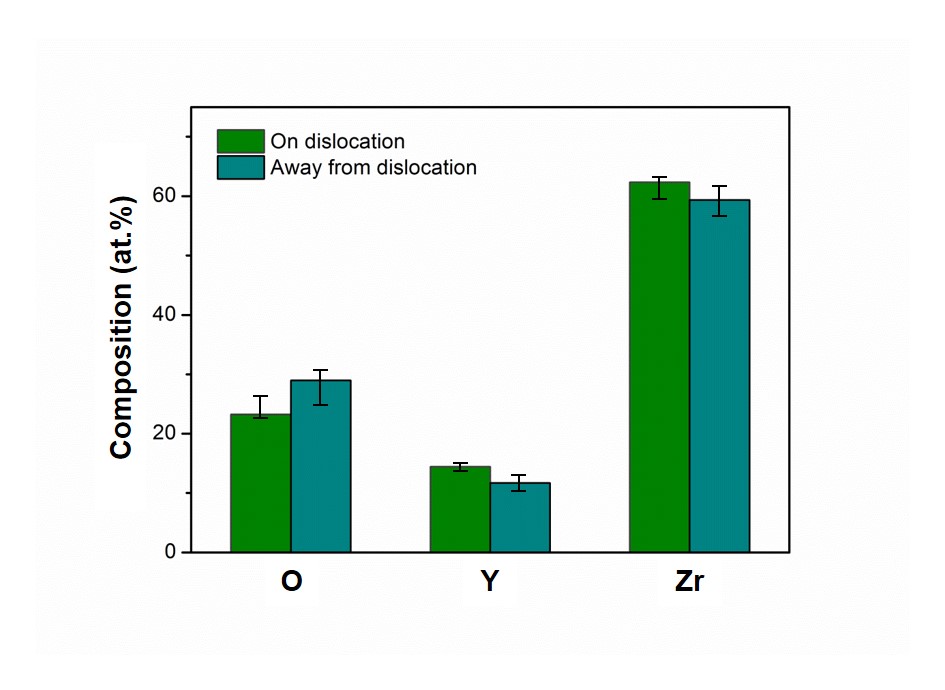}\
 
\textbf{Figure 6} Average atomic compositions \textit{on} dislocations and \textit{away from} dislocations in YSZ post-annealed at 1200 $^{o}$C by EDS. 
\\
\\
\textbf{Table 1} Comparison of the average atomic compositions \textit{on} dislocations and \textit{away from} dislocations in the as-compressed YSZ and the annealed one.

\includegraphics[width=140mm]{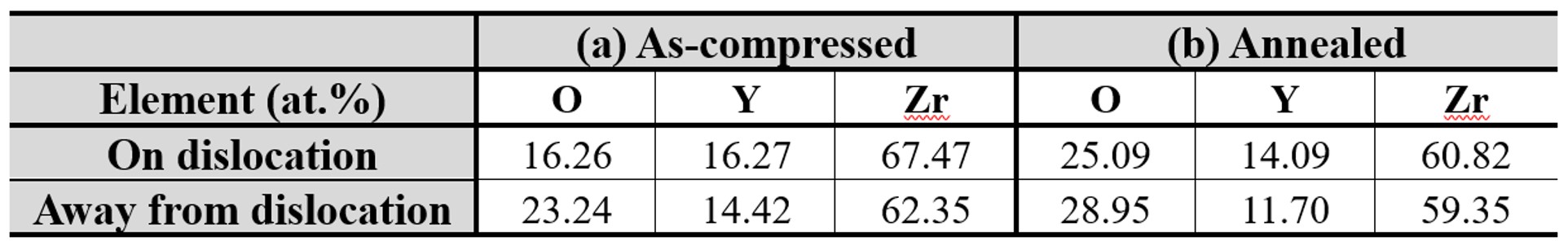}\
\section{Conclusions}
In summary, I first demonstrated oxygen depletion (oxygen vacancy accumulation) on dislocations in mechanically deformed YSZ with chemical characterization by EDS and EELS. Post-annealing at high temperature induced segregation of the acceptor-dopant (\textit{i.e.}, yittrium) to the dislocations, which is consistent with space charge layer model of grain boundary in YSZ. The result suggests that spatial aspects of oxygen vacancy and yttrium in space charge layers built in the vicinity of dislocations could be varied with temperature. 

\section{Acknowledgements}
This research was funded by the Department of Energy, Basic Energy Sciences under award DE SC0002633. This work made use of the Shared Experimental Facilities supported in part by the MRSEC Program of the National Science Foundation under award number DMR–1419807. I thank the Kwanjeong Educational Foundation for fellowship support.

\section{Reference}
[1] O. Yamamoto, Y. Arachi, H. Sakai, Y. Takeda, N. Imanishi, Y. Mizutani, M. Kawai, and Y. Nakamura. Zirconia based oxide ion conductors for solid oxide fuel cells. \textit{Ionics:4.403--408}. 1998.\

[2] J.A. Kilner. Fast oxygen transport in acceptor doped oxides. \textit{Solid State Ionics:129.13--23}. 2000.\

[3] F. Giannici, G. Gregori, C. Aliotta, A. Longo, J. Maier, and A. Martorana. Structure and Oxide Ion Conductivity: Local Order, Defect Interactions and Grain Boundary Effects in Acceptor-Doped Ceria. \textit{Chem. Mater.:26.5994--6006}. 2014.\

[4] M. Aoki, Y.-M. Chiang, I. Kosacki, L. J.-R. Lee, H. Tuller, and Y. Liu. Solute Segregation and Grain‐Boundary Impedance in High‐Purity Stabilized Zirconia. \textit{J. Am. Ceram. Soc.:79.1169--1180} 1996.\

[5] O. J. Durá, M. A. López de la Torre, L. Vázquez, J. Chaboy, R. Boada, A. Rivera-Calzada, J. Santamaria, and C. Leon. Ionic conductivity of nanocrystalline yttria-stabilized zirconia: Grain boundary and size effects. \textit{Phys. Rev. B:81.184301} 2010.\

[6] S. Kim, J. Fleig, and J. Maier. Space charge conduction: Simple analytical solutions for ionic and mixed conductors and application to nanocrystalline ceria. \textit{Chem. Phys.:6.2268--2273} 2003.\

[7] J. Maier. Ionic conduction in space charge regions. 
\textit{Prog. Solid State Chem.:23.171--263} 1995.\

[8] S. Kim, and J. Maier. On the Conductivity Mechanism of Nanocrystalline Ceria. \textit{J. Electrochem. Soc.:149.J73--J83} 2002.\

[9] R.A. De Souza, M. J. Pietrowski, U. Anselmi-Tamburini, S. Kim, Z. A. Munir, and M. Martin. Oxygen diffusion in nanocrystalline yttria-stabilized zirconia: the effect of grain boundaries. \textit{Phys. Chem. Chem. Phys.:10.2067--2072} 2008.\

[10] X. Guo, and R. Waser. Electrical properties of the grain boundaries of oxygen ion conductors: Acceptor-doped zirconia and ceria. \textit{Prog. Mater. Sci.:51.151--210} 2006.\

[11] K. Otsuka, A. Kuwabara, A. Nakamura, T. Yamamoto, K. Matsunaga, and Yuichi Ikuhara. Dislocation-enhanced ionic conductivity of yttria-stabilized zirconia. \textit{Appl. Phys. Lett.:82.877} 2003.\

[12] K.K. Adepalli, M. Kelsch, R. Merkle, and J. Maier. Influence of line defects on the electrical properties of single crystal TiO$_{2}$. \textit{Adv. Funct. Mater.:23.1798--1806} 2013.\ 

[13] D. Pergolesi, E. Fabbri, S.N. Cook, V. Roddatis, E. Traversa, and J.A. Kilner. Tensile Lattice Distortion Does Not Affect Oxygen Transport in Yttria-Stabilized Zirconia–CeO$_{2}$ Heterointerfaces. \textit{ACS Nano:6.10524--10534} 2012.\

[14] K. Song, H. Schmid, V. Srot, E. Gilardi, G. Gregori, K. Du, J. Maier, and P.A. van Aken. Cerium reduction at the interface between ceria and yttria-stabilised zirconia and implications for interfacial oxygen non-stoichiometry. \textit{APL Mater.:2.032104} 2014.\ 

[15] T.X.T. Sayle, S.C. Parker, and D.C. Sayle. Ionic conductivity in nano-scale CeO$_{2}$/YSZ heterolayers. \textit{J. Mater. Chem.:16.1067--1081} 2006.\

[16] S.R. Bishop, D. Marrocchelli, C. Chatzichristodoulou, N.H. Perry, M.B. Mogensen, H.L. Tuller, and E.D. Wachsman. Chemical Expansion: Implications for Electrochemical Energy Storage and Conversion Devices. \textit{Ann. Rev. Mater. Res.:44.205--239} 2014.\ 

[17] C. Chatzichristodoulou, P. V. Hendriksen and A. Hagen. Defect Chemistry and Thermomechanical Properties of Ce$_{0.8}$Pr$_{x}$Tb$_{0.2-x}$O$_{2-\delta}$. \textit{J. Electrochem. Soc.:157.B299--B307} 2010.\

[18] N. Q. Minh. Ceramic Fuel Cells. \textit{J. Am. Ceram. Soc.:76.563--588} 1993.\

[19] J. An, J.S. Park, A.L. Koh, H.B. Lee, H.J. Jung, J. Schoonman, R. Sinclair, T. M. Gür, and F.B. Prinz. Atomic Scale Verification of Oxide-Ion Vacancy Distribution near a Single Grain Boundary in YSZ. \textit{Sci. Rep.:3.2680} 2013.\

[20] X. Guo, and J. Maier. Grain Boundary Blocking Effect in Zirconia: A Schottky Barrier Analysis. \textit{J. Electrochem. Soc.:148.E121} 2001.\

[21] M.B. Ricoult, M. Badding, and Y. Thibault, 107th Annual Meeting, Exposition, and Technology Fair of the American Ceramic Society, 10-13 April (2005).\

[22] K. Matsui, H. Yoshida, and Y. Ikuhara. Grain-boundary structure and microstructure development mechanism in 2–8 mol$\%$ yttria-stabilized zirconia polycrystals. \textit{Acta Mater.:56.1315--1325} 2008.\

[23] Y. Lei, Y. Ito, N. D. Browning, and T. J. Mazanec. Segregation Effects at Grain Boundaries in Fluorite‐Structured Ceramics. \textit{J. Am. Ceram. Soc.:85.2359--2363} 2002.\

[24] M. Gerstl, T. Frömling, A. Schintlmeister, H. Hutter, and J. Fleig. Measurement of 18O tracer diffusion coefficients in thin yttria stabilized zirconia films. \textit{Solid State Ionics:184.23--26} 2011.\

[25] S. Weber, S. Scherrer, H. Scherrer, M. Kilo, M.A. Taylor, and G. Borchardt. SIMS analysis of multi-diffusion profiles of lanthanides in stabilized zirconias. \textit{Appl. Surface Sci.:203-204.656--659} 2003.\ 

\end{document}